\documentclass[10pt]{article}
\usepackage{amssymb}
\usepackage{amsfonts}
\usepackage{amsmath}
\usepackage{hyperref}

\makeatletter \@addtoreset{equation}{section} \makeatother

\newtheorem{theorem}{Theorem}
\newtheorem{lemma}{Lemma}

\newtheorem{remark}{Remark}

\newtheorem{proposition}{Proposition}

\begin{document}

\title{On fluctuations of eigenvalues of random band matrices}
\author{ M. Shcherbina
\\
Institute for Low Temperature Physics, Ukr. Ac. Sci \\
 47 Lenin ave, 61135 Kharkov Ukraine}
\date{}

\maketitle

\begin{abstract} We consider the fluctuation of linear eigenvalue statistics of random band $n\times n$ matrices
whose entries have the form  $\mathcal{M}_{ij}=b^{-1/2}u^{1/2}(|i-j|)\tilde w_{ij}$ with i.i.d. $w_{ij}$
possessing the $(4+\varepsilon)$th moment, where the function $u$
has a finite support $[-C^*,C^*]$, so that $M$ has  only
 $2C_*b+1$ nonzero  diagonals. The parameter $b$ (called the bandwidth) is assumed to grow with $n$ in a
  way that $b/n\to 0$. Without any additional assumptions on the growth of $b$ we prove
 CLT for linear eigenvalue statistics for a rather wide class of test functions.
 Thus we improve and generalize the results of the previous papers [8] and [11], where CLT was proven under the assumption
 $n>>b>>n^{1/2}$.  Moreover, we develop a  method which allows  to prove automatically the CLT  for linear
 eigenvalue statistics of the smooth test functions
 for almost all classical models of random matrix theory:
deformed Wigner and sample covariance matrices, sparse matrices, diluted random matrices, matrices with heavy tales etc.
\end{abstract}

%\textit{Key words:} random matrix, Wigner matrix, sample covariance matrix, Central Limit Theorem.
%
%\textit{Mathematical Subject Classification:} 15A52 (primary); 15A57 (secondary).

\section{Introduction and main results}\label{s:1}

Consider an ensemble of random symmetric $n\times n$ matrices with entries of the form
\begin{align}\label{band}&\mathcal{M}_{ij}=(u_{ij}/b)^{1/2}\tilde w_{ij},\quad u_{ij}=u(|i-j|/b)
\end{align}
where $\{\tilde w_{ij}\}_{ij}$ are i.i.d. (up to the symmetry $\tilde w_{ij}=\tilde w_{ji}$)  random variables,
satisfying the moment conditions
\begin{align}\label{band1}
 E\{\tilde w_{ij}\}=0,\quad  E\{|\tilde w_{ij}|^2\}=1,\quad E\{|\tilde w_{ij}|^4\}=3+\kappa_4,
\quad E\{|\tilde w_{ij}|^{4+\varepsilon}\}\le C<\infty,
%\notag
\end{align}
and $u(x)$ is a piece-wise continuous (with a finite number of jumps)
 continuous at $x=0$ function with a compact support, satisfying the conditions
\begin{align}\label{cond_u}
u(x)=u(-x),\quad 0\le u(x)\le C,\quad\int u(x)dx=1,\quad \mathrm{supp\,}u\subset[-C^*,C^*].
\end{align}
It is easy to see that the entries of $\mathcal{M}$ are nonzero only inside the band $|i-j|\le C^*b$. Hence for
fixed $b$ we have a matrix with a finite numbers of diagonals, while if $b\sim n$, we obtain some kind of the Wigner matrix,
with all of the entries  having the variances of  the same order (see \cite{W}). The model is now widely discussed
in mathematical literatures, since by non rigorous conjecture of \cite{FM:91} it is expected that the behavior of
local eigenvalue statistics demonstrates a kind of phase transition: for $b<<n^{1/2}$ the statistics is
of Poisson type and for $b>>n^{1/2}$ it is of the same type as for Wigner matrices. Till now this result is not
proven rigorously, but the problem is one of the most challenging in the random matrix theory (see, e.g. \cite{Sp},
\cite{EK:11}, \cite{EKY:13}, \cite{TS:12} and references therein).

It was proved many years ago
(see \cite{MKP:92}) that in the limit
\begin{align}\label{lim}
b\to\infty,\quad b/n\to 0,\quad \hbox{as} \quad n\to\infty,
\end{align}
the normalized eigenvalue counting measure converges weakly to the Wigner semicircle low, which has the density
\begin{align}\label{sc}
\rho_{sc}(\lambda)=\frac{1}{2\pi}\sqrt{4-\lambda^2}\mathbf{1}_{[-2,2]}.
\end{align}
This means that if we denote $\{\lambda_i\}_{i=1}^n$ the eigenvalues of $\mathcal{M}$, choose any bounded
integrable test function $\varphi$, and consider the linear eigenvalue statistics
of  the form
\begin{equation}\label{Linst}
\mathcal{N}_n[\varphi]=\sum_{j=1}^n\varphi(\lambda_j),\quad \mathcal{N}_n^\circ[\varphi]=
\mathcal{N}_n[\varphi]-E\{\mathcal{N}_n[\varphi]\},
\end{equation}
then in the limit (\ref{lim}) we have
\[E\{n^{-1}\mathcal{N}_n[\varphi]\}\to\int\varphi(\lambda)\rho_{sc}(\lambda)d\lambda,
\quad \mathrm{Var}\{\mathcal{N}_n[\varphi]\}\to 0.
\]
In particular, for  $\varphi(\lambda)=(\lambda-z)^{-1}$
\begin{align}\notag
&n^{-1}\mathcal{N}_n[\varphi]=n^{-1}\mathrm{Tr\,}(\mathcal{M}-z)^{-1}\to g(z),\\
& g(z)=\frac{1}{2}(-z+\sqrt{z^2-4}).
\label{g}\end{align}

The next natural question is the behavior of the fluctuations $\mathcal{N}_n^\circ[\varphi]$ in the same limit,
in particular, the behavior of its variance. This question was solved partially in the paper \cite{KK:02},
where the main term of the covariance of the traces of two resolvents was found in the case of Gaussian $w_{ij}$ and
under the additional restriction $b=n^\theta$, $1/3<\theta<1$. The next step was done in the papers \cite{Li-So:13} and
\cite{JSSo:14}, where  the Central Limit Theorem (CLT) for the random variable $\sqrt{b/n}\mathcal{N}_n^\circ[\varphi]$
was proved for sufficiently smooth test functions, but again under the technical condition
$n>>b>>n^{1/2}$.

The main result of the present paper is the proof of CLT for the linear eigenvalue statistics (\ref{Linst}) of the band matrices
under the limiting transition (\ref{lim}) without any additional restriction on the growth of $b$.

We consider the test functions from the  space $\mathcal{H}_s$, possessing the norm
\begin{equation}\label{norm}
    ||\varphi||_s^2=\int(1+2|k|)^{2s}|\widehat\varphi(k)|^2dk,\quad s>2,\quad
    \widehat\varphi(k)=\int e^{ikx}\varphi(x)dx.
\end{equation}

\begin{theorem}\label{t:band} Consider the model of band matrices (\ref{band})-(\ref{cond_u}) and any test function
possessing the norm (\ref{norm}) with $s>2$. Then  the sequence of random variables
$\sqrt{b/n}\mathcal{N}_n^\circ[\varphi]$ with  $\mathcal{N}_n^\circ[\varphi]$ of (\ref{Linst}) converges in distribution
in the limit (\ref{lim}) to the normal random variable with zero mean and the variance
\begin{align}\label{var}
V[\varphi]=&\frac{1}{\pi^2}\int_{0}^\pi dxdy\varphi(2\cos x)\varphi(2\cos y)
\int\frac{\partial^2}{\partial x\partial y}\log\bigg|\frac{1-\hat u(k)e^{i(x+y)}}{1-\hat u(k)e^{i(x-y)}} \bigg|dk\\
&+\frac{(u,u)\kappa _{4}}{\pi ^{2}}\left(
\int_{0}^\pi\varphi (2\cos x )\cos2x%
dx \right) ^{2}
+\frac{u(0)}{2\pi ^{2}}\left(
\int_{0}^\pi \varphi (2\cos x  )\cos x %
dx \right) ^{2},
\notag\end{align}
where $(u,u)=\int u^2(x)dx$ and $\hat u(k)$ is the Fourier transform of the function $u$ defined as in (\ref{norm})
\end{theorem}

To prove CLT for the band matrices, we use the  CLT for martingale (see \cite[Theorem 35.12]{Bi:95}).

\begin{theorem}\label{t:mart} Let $ X_{n,k}=E_{<k}\{Y-E_{k}Y\}$ be a martingale  differences array with respect to
independent random vectors $V_1,\dots,V_n$, $S_n=\sum_{k=1}^nX_k$, $\sigma_n=\sum_{k=1}^nE\{X_k^2\}=O(1)$. Assume that
\begin{align}\label{cond}
%&(1)\quad |X_k|\le C, k\le n;
&(1)\quad  \sum E\{X_k^4\}\le \varepsilon_n,\qquad
%&(3)\quad E_k\{X_k^4\}\le \varepsilon_n a_k,% \varepsilon_n\to 0,
(2)\quad \mathrm{Var}\Big\{\sum_{k=1}^nX_k^2\Big\}\le \tilde\varepsilon_n\,
%\notag
\end{align}
Then
\begin{align}\label{tm.1}
|E\{e^{itS_n}\}-e^{-t^2\sigma_n/2}|\le C'(t)(\varepsilon_n^{1/2}+\tilde\varepsilon_n^{1/2}).
\end{align}
\end{theorem}
\textit{Remark:} Here we have replaced a more general condition $\sum E\{X_k^21_{|X_k|>\delta}\}\to 0$ used in \cite{Bi:95}
by condition (1) which is more easy to check for the random matrix models.

The idea to use Theorem \ref{t:mart} for the proof of CLT in the random matrix theory is not new. Since the paper
of \cite{Ba-Si:04} it was used many times (see, e.g., \cite{Gu:13}, \cite{ORSo:14} and \cite{NY:14}),
but the method of the proof of CLT
used in the present paper allows to prove CLT  by the same way for all classical models of random matrix theory:
deformed Wigner  and sample covariance matrices, sparse and diluted random matrices etc. It becomes even simpler than
that for  band matrices, since the proof of condition (2) becomes simpler.

The paper is organized as follows. In Section 2.1 we give the sketch of the proof of CLT, introduce truncated band
matrix and explain how one can extend CLT from some special class of the test functions to all functions of $\mathcal{H}_s$.
In Section 2.2 we check conditions (\ref{cond}) and in Section 2.3 prove Lemma \ref{l:2}  (given in Section 2.1) about
the difference of linear eigenvalue statistics of initial and truncated matrices. In Section 3 we compute the variance
(\ref{var}). And in Section 4 the proofs of some auxiliary results (partially known before) are given in order to make the proof
of Theorem \ref{t:band} more self consistent.

\section{Proof of CLT}
\subsection{Strategy of the proof}
We start from the proof of CLT for  the truncated and "periodically continued" model:
\begin{align}\label{M_tr}
 &{M}_{ij}=(u_{ij}/b)^{1/2}w_{ij}, \qquad u_{ij}=u(|i-j|_n/b)\\
 & w_{ij}=\left\{\begin{array}{ll}\tilde w_{ij}1_{|\tilde w_{ij}|\le b^{1/2}}-
E\{\tilde w_{ij}1_{|\tilde w_{ij}|\le b^{1/2}}\},&|i-j|\le C^*b\\
\omega_{ij},&||i-j|-n|\le C^*b\end{array}\right.
\notag\end{align}
Here and below
\begin{align}\label{dist}|i-j|_n:=\max\{|i-j|,||i-j|-n|\},\end{align}
and $\{\omega_{ij}\}_{||i-j|-n|\le C^*b}$ are independent (up to the symmetry conditions) and independent from $\mathcal{M}$
copies of $w_{12}$. Thus we not only truncated the entries of $\mathcal{M}$, but also add entries in upper right and lower left
parts of it, in order to obtain the periodic distribution, i.e., invariant with respect to the shift $i\to |i+1|_n$.

Then the standard argument  gives us that for $|i-j|_n\le C^*b$
\begin{align}\label{M_tr1}
&E\{ w_{ij}\}=0,\quad E\{ |w_{ij}|^2\}=1+ O(b^{-1-\varepsilon/2}),\\
&E\{ |w_{ij}|^4\}=3+\kappa_4+O(b^{-\varepsilon/2}),\quad E\{ |\omega_{ij}|^8\}\le Cb^{4-\varepsilon/2}.
\notag\end{align}
Moreover, it is easy to see that
\begin{align}\label{M_tr2}
&n^{-1}E\big\{\mathrm{Tr\,}(\mathcal{M}-M)^2\big\}\le Cb^{-\varepsilon/2}.
\end{align}
Then, using Theorem \ref{t:mart}, we prove CLT for $\nu_{1n}:=(b/n)^{1/2}\mathcal{N}_n^\circ[\varphi_\eta,M]$
with the test functions of the form
\begin{equation}\label{phi_eta}
\varphi_\eta=\varphi*\mathcal{P}_\eta,
\end{equation}
where $*$ means a convolution, $\mathcal{P}_\eta$ is a Poisson kernel
\begin{equation}\label{P}
\mathcal{P}_\eta(\lambda)=\frac{\pi^{-1}\eta}{\lambda^2+\eta^2},
\end{equation}
and $\varphi\in\mathcal{H}_s\cap L_1(\mathbb{R})$.
It is easy to see that then
\begin{equation}\label{N(phi)}
\mathcal{N}_n[\varphi_\eta,M]=\pi^{-1}\int\varphi(\lambda)\Im\gamma_n(\lambda+i\eta)d\lambda.
\end{equation}

Then we shall prove the lemma
\begin{lemma}\label{l:2} Set $\mathcal{G}(z)=(\mathcal{M}-z)^{-1}$, $\tilde\gamma_n(z):=\mathrm{Tr\;}\mathcal{G}(z)$.
Then for any $z:\Im z>\eta$
\begin{align}\label{l2}
&\frac{b}{n}\mathrm{Var}\Big\{\gamma_n(z)-\tilde\gamma_n(z)\Big\}
\le Cb^{-\varepsilon/2}.
\end{align}
\end{lemma}
The lemma implies that for any $\varphi\in\mathcal{H}_s\cap L_1(\mathbb{R})$ if we set
$\nu_{2n}:=(b/n)^{1/2}\mathcal{N}_n^\circ[\varphi_\eta,\mathcal{M}]$, then
\begin{align*}\mathrm{Var}\{\nu_{2n}-\nu_{1n}\}=
&\frac{b}{n}\mathrm{Var}\Big\{\mathcal{N}_n[\varphi_\eta,\mathcal{M}]-\mathcal{N}_n[\varphi_\eta,{M}]\Big\}
\\&=\frac{b}{\pi^{2}n}
\int\int d\lambda_1d\lambda_2\varphi(\lambda_1)\varphi(\lambda_2)\\
&\times\mathrm{Cov}\{\Im\gamma_n(\lambda_1+i\eta)-\Im\tilde\gamma_n(\lambda_1+i\eta),
\Im\gamma_n(\lambda_2+i\eta)-\Im\tilde\gamma_n(\lambda_2+i\eta)\}\\&
\le Cb^{-\varepsilon}\int\int d\lambda_1d\lambda_2|\varphi(\lambda_1)\varphi(\lambda_2)|
\le  C'b^{-\varepsilon/2}.
\notag\end{align*}
Hence, for any fixed $x\in\mathbb{R}$
\begin{align}\label{CLT}
|E\{e^{ix\nu_{1n}}\}-E\{e^{ix\nu_{2n}}\}|\le x\mathrm{Var}^{1/2}\{\nu_{1n}-\nu_{2n}\}\le xCb^{-\varepsilon/4}.
\notag\end{align}
Thus,  CLT for $v_{1n}$ and Lemma \ref{l:2} imply CLT for $v_{2n}$, if the test function has the form
(\ref{phi_eta}).

To extend CLT to the test functions from $\mathcal{H}_s$, we use a  proposition (see \cite[Proposition
3.2.9]{PS:11}).
\begin{proposition} \label{p:CLTcont}
Let $\{\xi_{l}^{(n)}\}_{l=1}^{n}$ be a triangular array of random variables,
$\displaystyle\mathcal{N}_{n}[\varphi ]=\sum_{l=1}^{n}\varphi
(\xi _{l}^{(n)})$ be its linear statistics,
corresponding to a test function $\varphi :\mathbb{R}\rightarrow
\mathbb{R}$, and $\{d_n\}$ is some sequence of positive numbers.
%\[V_{n}[\varphi]=\mathrm{Var}\{\mathcal{N}_{n}[\varphi ]\}\]
%be the variance of $\mathcal{N}_{n}[\varphi ]$.
Assume that

(a) there exists a vector space $\mathcal{L}$ endowed with a norm $\|...\|$
and such that $V_{n}$ is defined on $\mathcal{L}$ and admits the bound
\begin{equation}
d_n\mathrm{Var}\{\mathcal{N}_{n}[\varphi ]\}\le C||\varphi ||^2,\;\forall \varphi \in \mathcal{L};
\label{b_var}
\end{equation}%
(b) there exists a dense linear manifold $\mathcal{L}_{1}\subset \mathcal{L}$ such that  CLT is valid for
$\mathcal{N}_{n}[\varphi ],\;\varphi \in \mathcal{L}_{1}$, i.e.,  there exists a continuous
quadratic functional $V:\mathcal{L}_{1}\rightarrow
\mathbb{R}_{+}$ such that we have uniformly in $x$, varying on any compact interval
\begin{equation}\label{limZ}
\lim_{n\rightarrow \infty }Z_{n}[x\varphi ]=e^{-x^{2}V[\varphi
]/2},\;\forall \varphi \in \mathcal{L}_{1},\quad where\quad Z_{n}[x\varphi ]:={E}\big\{
e^{ix d_n^{1/2}\mathcal{N}^\circ_{n}[\varphi ]}\big\}.
\end{equation}
Then  $V$ admits a continuous extension to $\mathcal{L}$ and CLT is valid for all $\mathcal{N}_{n}[\varphi ]$,
 $\varphi \in \mathcal{L}$.
\end{proposition}

The proposition allows to extend CLT from any dense subset of $\mathcal{H}_s$ for which we are able to prove CLT
on the whole $\mathcal{H}_s$, if we can
 check (\ref{b_var}). This can be done by using the another proposition (proven in \cite{S:11} and also \cite{ST:12})
 and Lemma \ref{l:b_var}.
\begin{proposition}\label{p:joh}
For any $s>0$ and any $\mathcal{M}$
\begin{equation}\label{pj.1}
\mathrm{Var}\{\mathcal{N}_n[\varphi,\mathcal{M}]\}\le C_s||\varphi||_s^2\int_0^\infty dy
e^{-y}y^{2s-1}\int_{-\infty}^\infty\mathrm{Var}\{\mathrm{Tr\,}\mathcal{G}(x+iy)\}dx.
\end{equation}
\end{proposition}
\begin{lemma}\label{l:b_var} If the conditions  (\ref{band}) and (\ref{cond_u})  are satisfied, then for any $0<y<1$
\begin{align}\label{b_v.1}
    \frac{b}{n}\int dx\mathrm{Var}\{\mathrm{Tr\,}\mathcal{G}(x+iy)\}\le Cy^{-4}\log y^{-1}
\end{align}
\end{lemma}
The proof of the lemma is given in Section 4.

Combining the proposition with  (\ref{b_v.1}), we prove (\ref{b_var}).

\subsection{Checking  of conditions (\ref{cond})} To apply Theorem \ref{t:mart}, we denote  $E_{p}$ the averaging
with respect to the variable $\{w_{p,j}\}_{j\ge p}$,  $E_{<p}=E_{1}\dots E_{p-1}$ and consider
\begin{align}\label{def_X}
X_p[\varphi_\eta]=\pi^{-1}(b/n)^{1/2}\int\varphi(\lambda)\Im \tilde X_p[\lambda+i\eta]d\lambda,\\
\tilde X_p[z]=E_{<p}\{\gamma_n(z)-E_{p}\gamma_n(z)\}.
\notag\end{align}
Then, according to Theorem \ref{t:mart}, we have to check condition (1)-(2) of (\ref{cond}) for $\{X_p[\varphi_\eta]\}$.
It is evident, that conditions (1) follow from the bounds
\begin{align}\label{cond.1}
%&(1)\quad|\tilde X_p[z]|\le 2/|\Im z|,&(2)\quad E_{p}\{|\tilde X_p[z]|^2\}\le Cb^{-1}\\
& E_{p}\{|\tilde X_p[z]|^2\}\le Cb^{-1},\quad |\tilde X_p[z]|\le C,
\end{align}
valid uniformly in $|\Im z|\ge\eta$. And since
\[
\mathrm{Var}\Big\{\sum X_p^2-\sum E_{p}\{X_p^2\}\Big\}=\sum_pE\big\{|X_p^2- E_{p}\{X_p^2\}|^2\big\}
\le\sum_pE\{X_p^4\}\le \varepsilon_n,
\]
condition (2) of (\ref{cond}) follows from
the uniform in $|\Im z_1|,|\Im z_2|\ge\eta$
bound
\begin{align}\label{cond.2}
& \mathrm{Var}\{\Sigma(z_1,z_2)\}\le \tilde\varepsilon_n,\\
&\Sigma(z_1,z_2):=\frac{b}{n}\sum_pE_{p}\{\tilde X_p[z_1]\tilde X_p[z_2]\}.
\notag\end{align}
 Let us prove (\ref{cond.1}) and (\ref{cond.2}).

Denote $M^{(p)}$ the $(n-1)\times(n-1)$ matrix which is obtained from $M$ by removing the $p$th line and column. Set also
\begin{equation}\label{G^1}
    G^{(p)}=(M^{(p)}-z)^{-1},\quad v^{(p)}:=(v_{p1},\dots,v_{pn})\in\mathbb{R}^{n-1},
    \quad v_{ij}:=u_{ij}^{1/2}w_{ij}.
\end{equation}
Use the  identities
\begin{align}\label{repr}
&G_{pp}=-A^{-1}_p,\;\quad {G}_{ij}=G^{(p)}_{ij}-Q^{(p)}_{ij},\quad\hbox{Tr }G-\hbox{Tr }G^{(p)}
=-\frac{\partial}{\partial z}\log A_p(z),\end{align}
where
\begin{align}\label{A,Q}
&A_p:= z+b^{-1/2}v_{pp}+b^{-1}(G^{(p)}v^{(p)},v^{(p)}),\\
& Q^{(p)}_{ij}=
b^{-1}A_p^{-1}({G}^{(p)}v^{(p)})_i({G}^{(p)}v^{(p)})_j.
\notag\end{align}
Since for the resolvent $G(z)=(M-z)^{-1}$ of any symmetric or hermitian  matrix $M$ and any vector  $m$
\begin{align}\label{sp_rel}
\Im(G(z) m,m)=\Im z(G(z) m,G(z) m),
\end{align}
we have for $|\Im z|\ge \eta$
\begin{align}\label{b_A_p}
&|A_p(z)|\ge|\Im A_p(z)|=|\Im z|\big(1+b^{-1}(G^{(p)}v^{(p)},G^{(p)}v^{(p)})\big)\ge\eta,\\
 & |\bar A_p|\ge |\Im \bar A_p|\ge\eta,\quad\mathrm{where} \quad\bar A_p:=  E_{p}\{A_p\},
\notag\\
&\|Q^{(p)}\|\le |A_p|^{-1}|b^{-1}(G^{(p)}v^{(p)},G^{(p)}v^{(p)})|\le \eta
\label{|Q|}\end{align}
and
\begin{align*}
\Big|\frac{ A_p'(z)}{A_p}\Big|\le
\frac{|1+b^{-1}((G^{(p)})^2v^{(p)},v^{(p)})|}{\Im A_p}\le\eta^{-1}\quad\Rightarrow\quad
|\tilde X_p|\le 2\eta^{-1},
\end{align*}
which implies the second inequality of (\ref{cond.1}).

The last relation of (\ref{repr}) yields
\begin{align*}&E_p\{\tilde X_p(z_1)\tilde X_p(z_2)\}=:\frac{\partial^2}{\partial z_1\partial z_1}D_p(z_1,z_2)
\\&D_p(z_1,z_2):=E_p\Big\{
E_{<p}\big\{(\log A_p(z_1))^\circ_p\big\}E_{<p}\big\{(\log A_p(z_2))^\circ_p\big\}\Big\}.
\end{align*}
Here and below for any random variable $\xi$ we denote $\xi^\circ_p=\xi-E_p\{\xi\}$.

Since $D_p(z_1,z_2)$ is an analytic function on $z_1,z_2:|\Im z_1|,|\Im z_2|\ge\eta/2$, in order  to prove
the first bound of (\ref{cond.1}, it suffices
to prove that uniformly in $|\Im z|\ge\eta/2$
\begin{align*}
&E_p\{|E_{<p}\{(\log A_p(z))^\circ_p\}|^2\}\le\eta^{-2}E_p\{|E_{<p}\{A_p^\circ(z)\}|^2\}\le Cb^{-1}.
\end{align*}
Evidently
\begin{align*}
E_{<p}\{A_p^\circ(z)\}=
b^{-1/2}v_{pp}+b^{-1}\sum_{i,j>p,i\not=j}G^{(p)}_{ij}(z)v_{pi}v_{pj}+
b^{-1}\sum_{i>p}G^{(p)}_{ii}(z)(v_{pi}^2-u_{pi}).
\end{align*}
Hence, averaging with respect to $E_p$ and using (\ref{M_tr1}), we obtain  the first bound of (\ref{cond.1}).
Similarly one can  get the relation which we need below
\begin{align}\label{A^4}
E_p\{|E_{<p}\{A_p^\circ(z')\}|^4\}\le Cb^{-1-\varepsilon/2}.
\end{align}

We are left to check (\ref{cond.2}). Writing $A_p=\bar A_p+A_p^\circ$, expanding $\log A_p$ around $\bar A_p$, and
using (\ref{A^4}), we obtain
\begin{align}\label{Cov}
\Sigma(z_1,z_2)=&\frac{b}{n}\sum_p E_p\{X_p(z_1)X_p(z_2)\}=\frac{\partial^2}{\partial z_1\partial z_1}\tilde \Sigma(z_1,z_2)
\\\tilde \Sigma(z_1,z_2)=:&\frac{b}{n}\sum D_p(z_1,z_2)
=\frac{b}{n}\sum (\bar A_p(z_1)\bar A_p(z_2))^{-1} T_p(z_1,z_2)\notag\\
&+\frac{b}{n}\sum \big(O(E_p\{|A_p^\circ(z_1)|^3\})+
O(E_p\{|A_p^\circ(z_2)|^3\}\big)\notag\\
=&\frac{b}{n}\sum (\bar A_p(z_1)\bar A_p(z_2))^{-1} T_{p}(z_1,z_2)+O(b^{-\varepsilon/4}),
%=\frac{b}{n}\Sigma_1
\notag\end{align}
where
\begin{align}\label{T_p}
 T_{p}(z_1,z_2):=&E_p\big\{E_{<p}\{ A_p^\circ(z_1)\}E_{<p}\{A_p^\circ(z_2)\}\big\}\\
=&2b^{-2}\sum_{i,j>p} u_{pi}u_{pj}E_p\big\{E_{<p}\{G^{(p)}_{ij}(z_1)\}E_{<p}\{G_{ij}^{(p)}(z_1)\}\big\}\notag\\
&+\kappa_4b^{-2}\sum u_{pi}^2E_p\big\{E_{<p}\{G^{(p)}_{ii}(z_1)\}E_{<p}\{G^{(p)}_{ii}(z_2)\}\big\}+b^{-1}u_{pp}.
\notag\end{align}

\begin{lemma}\label{l:G_ii} Given $\eta>0$ there exists $\delta(\eta)>0$ such that uniformly in $z:|\Im z|>\eta$
\begin{align}\label{var_G}
&\mathrm{Var}\{ G_{jj}^{(p)}\}\le b^{-\delta},\quad E\{|G_{jj}^{(p)}-E\{G_{jj}\}|\}\le C_0^2b^{-1},\\
&E\{ |G_{jj}^{(p)}(z)-g(z)|\}\le C_0b^{-\delta},
\notag\end{align}
where $g(z)$ is defined by (\ref{g}).
\end{lemma}
The proof of the  lemma is given in Section 4.
\begin{remark}\label{r:1}
Below we will often use a simple observation. If for  some random variables $|R_k|\le C_k$,  $\sum_kC_k\le C$,
and $f_k:E\{|f_k-f^*_k|\}\le C_1b^{-\delta}$, where $f^*_k$ are some constants,
then we have with the same $C$ and $C_0$ of (\ref{var_G})
\begin{align}\label{obs}
\sum R_kf_k=\sum R_k f_{k}^*+r,\quad E\{|r|\}\le CC_1b^{-\delta}.
\end{align}
\end{remark}
In particular, since in view of (\ref{T_p}) $| T_{p}(z_1,z_2)|\le Cb^{-1}$, we have
\begin{align}\label{Cov1}
\tilde\Sigma(z_1,z_2)&=\frac{b}{n}\sum (\bar A_p(z_1)\bar A_p(z_2))^{-1} T_{p}(z_1,z_2)+o(1)\\&=
\frac{2}{bn}\sum g(z_1)g(z_2) T_{p}'(z_1,z_2)+\kappa_4(g(z_1)g(z_2))^{2}+u(0)g(z_1)g(z_2)+o(1),
\notag\end{align}
where $T_{p}'(z_1,z_2)$ is the first sum in the r.h.s. of (\ref{T_p}).
The  constant  term here does not contribute into the variance
of $\Sigma(z_1,z_2)$, so it is not important in the proof of (\ref{cond.2}).

 Let us denote $\tilde M^{(<p)}$ the matrix $M$ whose entries $w_{ij}$ with $\min\{i,j\}<p$ are replaced by
$\tilde w_{ij}$ which are independent from all $\{w_{kl}\}_{k,l=1}^n$ and have the same distribution as $w_{ij}$.
Let also $\tilde M^{(<p,q)}$ be the matrix $\tilde M^{(<p)}$ without $q$th line and column. We denote also
$\tilde E_{<p}$ the averaging with respect to all $w_{ij}$ and $\tilde w_{ij}$ with $\min\{i,j\}<p$. Set
\begin{align}\label{G^<i}
\tilde G^{(<p,q)}=(\tilde M^{(<p,q)}-z)^{-1},\quad \tilde G^{(<p)}=(\tilde M^{(<p)}-z)^{-1}.
\end{align}
Then evidently
\begin{align*}
T_{p}'(z_1,z_2)=\sum_{jk}E_p\{\tilde E_{<p}\{\tilde G^{(<p,p)}_{jk}(z_1)G^{(p)}_{jk}(z_2)\}\}u_{jp}u_{kp}
\\=E_p\{\tilde E_{<p}\{\mathrm{Tr} \tilde G^{(<p,p)}(z_1)I^{(p)}G^{(p)}(z_2)I^{(p)}\}\},
\end{align*}
where we denote by $I^{(p)}$ the diagonal matrix with the entries
\begin{align}\label{I^p}
I^{(p)}_{jk}=\delta_{j,k}u_{kp}\mathbf{1}_{k>p}.
\end{align}
Moreover, if we replace $G^{(p)}$ in (\ref{T_p}) by $G$ and set
\begin{align}\label{T''}T_{p}''(z_1,z_2)=&\sum_{i,j>p} u_{pi}u_{pj}E_p\{E_{<p}\{G_{ij}(z_1)\}E_{<p}\{G_{ij}(z_1)\}\}
\\=&E_p\{\tilde E_{<p}\{\mathrm{Tr\,} \tilde G^{(<p)}(z_1)I^{(p)}G(z_2)I^{(p)}\}\},
\notag\end{align}
then in view of (\ref{repr}) and (\ref{|Q|})
\begin{align}\label{de_T}
|T_{p}''(z_1,z_2)-T_{p}'(z_1,z_2)|\le &|E_p\{\tilde E_{<p}\{\mathrm{Tr\,} \tilde Q^{(p)}(z_1)I^{(p)}G(z_2)I^{(p)}\}\}|\\&+
|E_p\{\tilde E_{<p}\{\mathrm{Tr\,} \tilde G^{(<p,p)}(z_1)I^{(p)}Q^{(p)}I^{(p)}\}\}|\le C,
%&\le
%\frac{C}{bn}\sum_{\alpha,\alpha'=1,2}\sum_{i,j,p} |u_{pi}u_{pj}|\Big(E\{ |G_{ij}(z_\alpha)||Q^{(p)}_{ij}(z_{\alpha'})|\}
%+E\{ |Q^{(p)}_{ij}(z_{\alpha})||Q^{(p)}_{ij}(z_{\alpha'})|\}\Big)\notag\\&\le C(S^{1/2}+S)
\notag\end{align}
where we have used that since $Q^{(p)}$ is a rank one matrix with a bounded norm, we have for any bounded matrix $B$
\begin{align*}\mathrm{Tr\,}Q^{(p)}B\le \|B\|\|Q\|.
\end{align*}
Thus we need to study the variance of
\begin{align}\label{S_1}
\Sigma_1=&\frac{1}{bn}\sum_p T_{p}''(z_1,z_2).
\end{align}

To prove (\ref{cond.2}), it suffices to show that
\[\mathrm{Var}\{\Sigma_1\}=\sum_rE\{|E_{<r}^2\{\big(\Sigma_1\big)^\circ_r\}|^2\}\to 0.
%\quad\mathrm{where}\quad \Delta_r:=\big(\Sigma_1\big)^\circ_r.
\]
The last relation is a corollary of of the bounds, which we are going to prove
\begin{align}\label{varp}
n^2E\{|(\Sigma_1\big)^\circ_r|^2\}\le C,\quad r=1,\dots,n.
\end{align}
By (\ref{S_1}),
\begin{align}\label{var.1}
n\big(\Sigma_1\big)^\circ_r=\frac{1}{b}\sum_{p\le r}\big(T_p''(z_1,z_2)\big)^\circ_r.
\end{align}
Notice also that $\big(T_p''(z_1,z_2)\big)^\circ_r=0$ for $p\ge r+1$,
hence the  sum in (\ref{var.1a}) is over $p\le r$.

 Then (\ref{repr}) yields
\begin{align*}
\big(T_p''(z_1,z_2)\big)^\circ_r=&\Big(\tilde E_{\le p}\{\mathrm{Tr\,} \tilde G^{(<p)}(z_1)I^{(p)}G(z_2)I^{(p)}\}-
\tilde E_{\le p}\{\mathrm{Tr\,}\tilde G^{(<p,r)}(z_1)I^{(p)}G^{(r)}(z_2)I^{(p)}\}\Big)^\circ_r
\\
=&\Big(
\tilde E_{\le p}\big\{(A_rb)^{-1}(G^{(r)}(z_2)I^{(p)}\tilde G^{(<p,r)}(z_1)I^{(p)}G^{(r)}(z_2)v^{(r)},v^{(r)})\big\}\Big)^\circ_r\\
&+\mathrm{sim}+\Big(\tilde E_{\le p}\big\{(A_rb)^{-2}(G^{(r)}(z_2)I^{(p)}
\tilde G^{(r)}(z_1)\tilde v^{(r)},v^{(r)})^2\big\}\Big)^\circ_r\\
=:&\big(\tilde F^{(r)}_{1p}(z_1,z_2)\big)^\circ_r+\big(\tilde F^{(r)}_{1p}(z_2,z_1)\big)^\circ_r+
\big(\tilde F^{(r)}_{2p}(z_1,z_2)\big)^\circ_r,
\end{align*}
where  "+sim" means the  adding of the term which can be obtained from the previous one
by replacing $z_2$ and $z_1$. Since $E\{|\xi^\circ_r|^2\}\le E\{|\xi|^2\}$ for any random variable $\xi$,
(\ref{var.1}) yields
\begin{align}\label{var.1a}
n^2&E\{|\big(\Sigma_1\big)^\circ_r|^2\}\le CE\Big\{\Big|b^{-1}\sum_{p\le r}\big( \tilde F^{(r)}_{1p}(z_1,z_2)
+\tilde F^{(r)}_{1p}(z_2,z_1)+\tilde F^{(r)}_{2p}(z_1,z_2)\big)
\Big|^2\Big\}\\
 &\le CE\Big\{\Big|b^{-2}\sum_{p\le r}E_{\le p}\big\{\big(I^{(p)}G^{(r)}(z_1)v^{(r)},G^{(r)}(z_1)v^{(r)}\big)
\big(1+b^{-1}( v^{(r)}, v^{(r)})\big)\big\}\Big|^2\Big\}+\mathrm{ sim}\notag\\
&=: CE\Big\{\Big|b^{-2}\sum_{p\le r}\big(F^{(r)}_p(z_1)+F^{(r)}_p(z_2)\big)\Big|^2\Big\}.
\notag\end{align}

To sum in the r.h.s of (\ref{var.1a}) with respect to $p$ we would like to use the property
\begin{align}\label{sum_I}
\sum_{p=1}^nI^{(p)}\le CbI,
\end{align}
but since $p$ appears not only in $I^{(p)}$, we need to remove $p$ from the other places first.
Write
\begin{align*}
&E\big\{n^2\big|\big(\Sigma_1\big)^\circ_r\big|^2\big\}\le Cb^{-4}\sum_{p\le q\le r}E\{ F^{(r)}_pF^{(r)}_q\}
\\
\le&Cb^{-4}\sum_{ q=1}^r\sum_{p=1}^qE\Big\{E_{\le q}\big\{(I^{(p)}G^{(r)}v^{(r)},G^{(r)}v^{(r)})
(1+b^{-1}(v^{(r)},v^{(r)}))\big\}F^{(r)}_q\Big\}
\\
\le&Cb^{-2}\sum_{ q=1}^rE\Big\{E_{\le q}\big\{b^{-1}(v^{(r)},v^{(r)})(1+b^{-1}(v^{(r)},v^{(r)}))\big\}F^{(r)}_q\Big\}
\\
\le&Cb^{-2}\sum_{ q=1}^{r-C_*b}E\Big\{\big(I^{(q)}G^{(r)}v^{(r)},G^{(r)}v^{(r)}\big)\big(1+b^{-1}(v^{(r)},v^{(r)})\big)^3\Big\}
\\
&+CE\{(1+b^{-1}(v^{(r)},v^{(r)}))^4\}\le C'E\{(1+b^{-1}(v^{(r)},v^{(r)}))^4\}.
\end{align*}
Here in the first line we use (\ref{var.1a}), in the second line we  use first that for $p\le q$ the
averaging $E_{\le p}$ can be replaced by $E_{\le q}$, and  then
use (\ref{sum_I}) for summation
  over $p\le r$. The third line follows from the second one in view of the bound $\|G^{(r)}\|\le C$.
   Next we split the sum over $q$ into two parts: one over $q<r-C^*b$ and another over $r-C^*b\le q\le r$,
 and observed
that for the $q$ in the first part $(v^{(r)},v^{(r)})$ is a constant with respect to the averaging $E_{<q}$, hence
\begin{align*}
E\big\{E_{<q}&\big\{b^{-1}(v^{(r)},v^{(r)})\big(1+b^{-1}(v^{(r)},v^{(r)})\big)\big\}F^{(r)}_q\big\}
\\&=E\big\{\big((G^{(r)})^*I^{(q)}G^{(r)}v^{(r)},v^{(r)}\big)b^{-1}(v^{(r)},v^{(r)})
\big(1+b^{-1}(v^{(r)},v^{(r)})\big)^2\big\}.
\end{align*}
Then we can take the sum over $q<r-C^*b$, using again the bound (\ref{sum_I}), and finish to estimate the sum using
the bound $\|G^{(r)}\|\le C$.  As for the terms  with $r-C^*b\le q\le r$,  they are estimated just using
the boundedness of $\|G^{(r)}\|$ and $\|I^{(p)}\|$.
Thus we have proved (\ref{varp}).

$\square$

\medskip

\subsection{Proof of Lemma \ref{l:2}} Set
\begin{align*}\mathcal{G}^{(p)}:=(\mathcal{M}^{(p)}-z)^{-1},\quad
\mathcal{A}_p:=z+b^{-1}(\mathcal{G}^{(p)}\tilde v^{(p)},\tilde v^{(p)}),\quad \Delta A_p:=\mathcal{A}_p-A_p.
\end{align*}
The same argument as in the previous section implies that it suffices to check that
\begin{align}\label{l2.1a}
\frac{b}{n}\sum_{p}E\{ |\Delta A_p-E_p\{\Delta A_p\}|^2\}\to 0.
\end{align}
Since  we know that (see (\ref{T_p}))
\begin{align*}
\frac{b}{n}\sum_{|p|_n\le C^*b}E\{ |\Delta A_p-E_p\{\Delta A_p\}|^2\}
\le \frac{b}{n}\sum_{|p|_n\le C^*b}2\Big(E\{|A_p^\circ|^2\}
+E\{|\mathcal{A}_p^\circ|^2\}\Big)\le \frac{Cb}{n},\end{align*}
we conclude that it suffices to prove that
\begin{align}\label{l2.1}
\frac{b}{n}\sum_{|p|_n>C^*b}E\{ |\Delta A_p-E_p\{\Delta A_p\}|^2\}\to 0.
\end{align}
Let us write
\begin{align}\label{l2.2}
\Delta A_p=&b^{-1/2}\Delta v_{pp}+b^{-1}(\mathcal{G}^{(p)}\Delta v^{(p)},\Delta v^{(p)})
+2b^{-1}(\mathcal{G}^{(p)}\Delta v^{(p)}, v^{(p)})\\&
+ b^{-1}((\mathcal{G}^{(p)}-{G}^{(p)}) v^{(p)}, v^{(p)})=:J_{0p}+J_{1p}+2J_{2p}+J_{3p}.
\notag\end{align}
Averaging with respect to $ v^{(p)}$ and $\tilde v^{(p)}$ we get similarly to (\ref{T_p}) for $|p|_n\ge cb$
\begin{align}\label{l2.3}
E\{|J_{1p}-&E_p\{J_{1p}\}|^2\}=b^{-2}\sum_{i\not=j}E\{|\mathcal{G}^{(p)}_{ij}|^2(v_{pi}-\tilde v_{pi})^2
(v_{pj}-\tilde v_{pj})^2\}
\\&\qquad\qquad+b^{-2}\sum_{i}E\{|\mathcal{G}^{(p)}_{ii}|^2\}(v_{pi}-\tilde v_{pi})^4\}\notag\\
\le&b^{-4-\varepsilon}\sum_{i\not=j}E\{|\mathcal{G}^{(p)}_{ij}|^2\}I^{(p)}_{ii}I^{(p)}_{jj}+
b^{-2-\varepsilon/2}\sum_{i}E\{|\mathcal{G}^{(p)}_{ii}|^2\}I^{(p)}_{ii}\le Cb^{-1-\varepsilon/2}.
\notag\end{align}
Similarly
\begin{align}\label{l2.4}
E\{|J_{2p}-E_p\{J_{2p}\}|^2\}\le Cb^{-2-\varepsilon/2},\quad E\{|J_{0p}-E_p\{J_{0p}\}|^2\}\le Cb^{-2-\varepsilon/2}.
\end{align}
In addition, again similarly to (\ref{T_p}) we have
\begin{align}\label{l2.5}
E\{|J_{3p}-E_p\{J_{3p}\}|^2\}\le &Cb^{-2}E\{\mathrm{Tr\,}I^{(p)}(\mathcal{G}^{(p)}-{G}^{(p)})I^{(p)}
(\mathcal{G}^{(p)*}-{G}^{(p)*})\}.
\end{align}
Now by the same way as in (\ref{T''})-(\ref{de_T}) we can replace here $\mathcal{G}^{(p)}$ by $\mathcal{G}$
and ${G}^{(p)}$ by ${G}$ with an error $O(b^{-2})$:
\begin{align}\label{l2.6}
E\{|J_{3p}-E_p\{J_{3p}\}|^2\}\le &2b^{-2}E\{\mathrm{Tr\,}I^{(p)}(\mathcal{G}-{G})I^{(p)}
(\mathcal{G}^{*}-{G}^{*})\}+O(b^{-2}).
\end{align}
The resolvent identity implies
\[\mathcal{G}-{G}={G}(M^{(p)}-\mathcal{M})\mathcal{G}=-{G}\Delta M\mathcal{G}.
\]
Hence, the last term in the r.h.s. of (\ref{l2.5}) can be estimated as
\begin{align*}
b^{-2}E\{\mathrm{Tr\,}I^{(p)}(\mathcal{G}-{G})I^{(p)}(\mathcal{G}^{*}-{G}^{*})\}&=
b^{-2}E\{\mathrm{Tr\,}I^{(p)}{G}\Delta M\mathcal{G}I^{(p)}\mathcal{G}^{*}\Delta M{G}^{*})\}\\&\le
Cb^{-2}E\{\mathrm{Tr\,}I^{(p)}{G}(\Delta M)^2{G}^{*})\}.
\end{align*}
Hence, using (\ref{sum_I}) and (\ref{M_tr2}), we obtain
\begin{align}\label{l2.7}
\frac{b}{n}\sum_{C^*b< p<n- C^*b} E\{|J_{3p}-E_p\{J_{3p}\}|^2\}&\le
Cn^{-1}b^{-1}E\{\mathrm{Tr\,}{G}(\Delta M)^2{G}^{*}\}\\
&\le Cn^{-1}b^{-1}E\{\mathrm{Tr\,}(\Delta M)^2\}\le Cb^{-1-\varepsilon/2}.
\notag\end{align}
Combining (\ref{l2.7}) with (\ref{l2.2})-(\ref{l2.5}), we get (\ref{l2.1}).

$\square$

\section{Variance}
In view of (\ref{S_1}) to find $\Sigma_1$, it suffices to find the main order of
$b^{-1}E\{T_p''(z_1,z_2)\}$ defined in (\ref{T''}). For this aim it suffices to compute for any $i$ the main order
of
\[t_i=\sum_{j>p}u_{pj}\tilde E_{<p}\{\tilde G_{ij}(z_1)G_{ij}(z_2)\}.\]
Consider
\begin{align}\label{eq_1}
s_i:=&\sum_{j>p}u_{pj}\tilde E_{<p}\Big\{\tilde G_{ij}(z_1)\sum_{k}b^{-1/2}v_{ik}G_{kj}(z_2)\Big\}\\
=&\sum_{j>p}u_{pj}\tilde E_{<p}\Big\{\tilde G_{ij}(z_1)\sum_{k}\Big(b^{-1/2}v_{ik}-z_2\delta_{ik}
+z_2\delta_{ik}\Big)G_{kj}(z_2)\Big\}\notag\\
=&\sum_ju_{pj}\delta_{ij}E\{G_{ii}(z_1)\}+z_2t_i=u_{pi}g(z_1)+z_2t_i+O(b^{-\delta/2}),
\notag\end{align}
where we used Lemma \ref{l:G_ii} for the last equality.

The idea is to compute the l.h.s. above in  a way which gives us an equation with respect to $\{t_i\}_{i>p}$.
It is possible by using the formula (see e.g.\cite{PS:11}) valid for any random variable $\xi$ which has zero mean
and possesses
$m+2$ moments, and any function $F$, possessing $m+1$ bounded derivatives
\begin{align}\label{cum}
E\{\xi F(\xi)\}=\sum_{s=1}^{m}\frac{\kappa_{s+1}E\{ F^{(s)}(\xi)\}}{s!}+r_m,\quad |r_m|\le CE\{|\xi|^{m+2}\}\max|F^{(m+1)}|.
\end{align}
Applying this formula for $\xi=b^{-1/2}v_{ik}$, $m=4$, and $F_{ijk}=\tilde G_{ij}(z_1)G_{ik}(z_2)$, we get
\begin{align}\label{eq_2}
s_i=&-\sum_{j>p}u_{pj}\tilde E_{<p}\Big\{\tilde G_{ji}(z_1)G_{ij}(z_2)\sum_{k}b^{-1}u_{ik}G_{kk}(z_2)\Big\}\\&-
\sum_{j>p}b^{-1}u_{ik}u_{pj}\tilde E_{<p}\{\tilde G_{ii}(z_1)\tilde G_{jk}(z_1)G_{kj}(z_2)\sum_{k}G_{jk}(z_2)\}\notag
\\&+R_1+R_2+R_3+R_4.
\notag\end{align}
Here we  used the differentiation formula for the resolvent of any symmetric matrix $M$
\begin{align}\label{diff}
\frac{d}{d M_{ik}}G_{sl}(z)=-G_{sk}(z)G_{il}(z)-G_{si}(z)G_{kl}(z)
\end{align}
Two sums written in the r.h.s of (\ref{eq_2}) collect the terms,
corresponding to $s=1$ in the r.h.s. of (\ref{diff}). The remainder $R_1$ collects the terms, corresponding to $s=2$
 in the r.h.s. of (\ref{diff}). The remainders $R_2$ and $R_3$ collect the terms, corresponding to $s=3$
 and $s=4$  respectively. And the remainder $R_4$ appears because of the remainder
in (\ref{cum}). Let us analyze the order of each of these terms.
By (\ref{diff})
\begin{align*}R_1=&-b^{-1}\sum_{j>p,k}u_{jp}u_{ik}\tilde E_{<p}\{\tilde G_{ij}(z_1)G_{ik}(z_2)G_{kj}(z_2)\}\\&-
b^{-1}\sum_{j,k>p}u_{jp}u_{ik}\tilde E_{<p}\{\tilde G_{ik}(z_1)\tilde G_{ij}(z_1)G_{kj}(z_2)\}\\
=&-b^{-1}\tilde E_{<p}\{(\tilde G I^{(p)}G I^{(i,p)}G )_{ii}\}
-b^{-1}\tilde E_{<p}\{(\tilde G I^{(p)}G I^{(i,p)}\tilde G )_{ii}\}=O(b^{-1}).
\end{align*}
where $I^{(i,p)}_{lk}=\delta_{lk}u_{lk}1_{k>p}$.

To estimate $R_2$,  observe that by (\ref{diff})  after two differentiation we obtain the sum of terms
of the type $ \hat G_{l_1l_2}\hat G_{l_3l_4}\hat G_{l_5l_6}\hat G_{l_7,l_8}$, where
$\hat G$ can be $G$ or $\tilde G$ and the set of indexes  $l_1,l_2\dots l_7,l_8$ contains  3 times  $i$, 3 times
$k$, and 2 times $j$,
but $\hat G_{jj}$ can not appear. Thus, each term contains either $\hat G_{jk}\hat G_{ji}$ or $\hat G_{jk}\hat G_{jk}$,
 or $\hat G_{ji}\hat G_{ji}$. Any of this combinations
after  summation with respect to $j$  gives us $O(1)$. Hence, after summation with respect to $k$ we obtain
$O(b)$. But the factor which appears because of the third cumulant is $b^{-3/2}$, hence
$R_2=O(b^{-1/2})$.
By the same argument
$R_3=O(b^{-1})$.

Finally, to estimate $R_4$, observe that we have two summations with respect to $p<j<p+C_*b$ and $i-C_*b<k<i+C_*b$,
and the factor which appears because of $b^{-3}E\{|v_{ik}|^6\}$ is bounded by $b^{-2-\varepsilon/2}$.
At the last step of transformations of (\ref{eq_2}) we write
 \[G_{kk}(z_2)=g(z_2)+(G_{kk}(z_2)-g(z_2)),\quad G_{ii}(z_1)=g(z_1)+(G_{ii}(z_1)-g(z_1))\]
and  use the bound (\ref{var_G}). Then we obtain
\[ s_i=-g(z_2)t_i-g(z_1)\sum_{k} U_{ik}^{(p)}t_k+r_i,\quad r_i\le C b^{-\varepsilon/2},
\]
where
\begin{align}\label{U}
U_{ik}=b^{-1}u_{ik},\quad U^{(p)}_{ik}=b^{-1}u_{ik}\mathbf{1}_{i>p}\mathbf{1}_{k>p}.
\end{align}
Combining (\ref{eq_1}) and (\ref{eq_2}) with  above estimates for the reminders and using that
by (\ref{g}) we have $(z_2+g(z_2))=-g^{-1}(z_2)$, we obtain the system of equations
\begin{align}\label{eq_3}
&\big((\zeta-U^{(p)})t\big)_i=u^{(p)}_i+r'_i,\quad r_i'\le C (b^{-\varepsilon/2}+b^{-\delta/2}),\\
&\mathrm{with}\quad \zeta=(g(z_1)g(z_2))^{-1},\quad u^{(p)}_i=\mathbf{1}_{i>p}u_{pi}.
\notag\end{align}
Since $|g(z_1)g(z_2)|<1$ and
\[\|U^{(p)}\|\le \max_{k}\sum_i|U_{ki}|\le 1+o(1),\]
the operator $(\zeta-U^{(p)})^{-1}$ can be
defined by the Neumann series
\[(\zeta-U^{(p)})^{-1}=\sum_{m=0}^\infty \zeta^{-m-1}(U^{(p)})^m,\]
and it possesses the properties
\begin{align}\label{prU}
\sum_{k}|(U^{(p)}-\zeta)^{-1}_{ik}|\le C,\;\; i>p,\quad
|(U^m)_{ii}|\le Cb^{-1},\; 1\le i\le n.
\end{align}
Application of $(\zeta-U^{(p)})^{-1}$ to both parts of (\ref{eq_3}) and (\ref{prU}) imply
\begin{align}\label{eq_4}
t_i=&\big((U^{(p)}-\zeta)^{-1}u^{(p)}\big)_i+\tilde r_i,\quad |\tilde r_i|\le C (b^{-\varepsilon/2}+b^{-\delta/2}),\\
\Rightarrow& b^{-1}T_p''(z_1,z_2)=b^{-1}\big((\zeta-U^{(p)})^{-1}u^{(p)},u^{(p)}\big)+o(1),\notag\\
\Rightarrow&E\{\Sigma_1\}=\frac{2}{n\zeta}\sum_p b^{-1}\big((\zeta-U^{(p)})^{-1}u^{(p)},u^{(p)}\big)
+o(1),\notag
\end{align}
where $\Sigma_1$ was defined in (\ref{S_1}).
\begin{proposition}\label{p:U}
 Let the matrices $U$ and  $U^{(p)}$ be defined by (\ref{U}), where $\{u_{i,j}\}$ satisfy conditions
(\ref{cond_u}),  the vectors $u^{(p)}$ be defined by (\ref{eq_3}), and $|\zeta|>1$.
Then
\begin{align}\label{pU.1}
&\frac{1}{\zeta n}\sum_{p=1}^{n}b^{-1}((\zeta-U^{(p)})^{-1}u^{(p)},u^{(p)})
=-\frac{b}{n}\Big(\mathrm{Tr}\log(1-\zeta^{-1}U)+\zeta^{-1}\mathrm{Tr}U\Big)
 +O(b^{-1}).
\end{align}
\end{proposition}
\textit{Proof.} Denoting by $S_1(z)$
the l.h.s. of (\ref{pU.1}) and  by $S_2(z)$
the main term in the r.h.s. of (\ref{pU.1}), we have
\begin{align*}
S_2(z)=&\frac{b}{n}\sum_{m=2}^{\infty}m^{-1}\zeta^{-m}\sum U_{i_1i_2}\dots U_{i_mi_1}\\
=&\frac{b}{n}\sum_{p=1}^n\sum_{m=2}^{\infty}m^{-1}\zeta^{-m}\sum_{\min\{i_1,\dots,i_m\}=p} U_{i_1i_2}\dots U_{i_mi_1}\notag\\
=&\frac{b}{n}\sum_{p=1}^n\sum_{m=2}^{\infty}\zeta^{-m}\sum_{i_2,\dots,i_{m-1}>p} U_{pi_2}\dots U_{i_mp}+O(b^{-1})\notag\\
=&\frac{1}{bn}\sum_{p=1}^{n}\sum_{m=2}^{\infty}\zeta^{-m}((U^{(p)})^{m-2}u^{(p)},u^{(p)})+O(b^{-1})=S_1(z)+O(b^{-1}).
\notag\end{align*}
The term $O(b^{-1})$ appears in the third line above  as a sum of the terms, which have at least two $p$ among $\{i_1,\dots,i_m\}$.
But the contribution of these terms for fixed $m$ in view of (\ref{prU}) can be estimated as
\[m|z|^{-m-1}\sum_{k=1}^{m-1} (U^{k})_{pp}(U^{m-k})_{pp}\le m^2|z|^{-m-1}b^{-2}.\]
After summation with respect to $m$ and multiplication by $b$ we obtain $O(b^{-1})$.

$\square$

Now observe that  the r.h.s. of (\ref{pU.1}) has a limit, as $n,b\to\infty$ like in (\ref{lim}).
\begin{align*}
&S_2(z)=\sum_{m=2}^{\infty}\frac{1}{m\zeta^{m}}\int u(x_1-x_2)\dots u(x_{m-1}-x_{m})(u(x_{m})-u(x_1))d\bar x
+r_{n,b}\\&=-\int \log\big(1-\zeta^{-1}\hat u(k)\big)dk-\zeta^{-1}u(0)+o(1),
%\\&  \hat u(k)=\int u(x)e^{ikx} dx
\end{align*}
where $\hat u$ is the Fourier transform of the function $u$ defined as in (\ref{norm}).
Hence, the proposition and the last line of (\ref{eq_4}) yield
\begin{align*}
E\{\Sigma_1\}=-2\Big(\int \log\big(1-\zeta^{-1}\hat u(k)\big)dk+\zeta^{-1}u(0)\Big)+o(1).
\end{align*}
Thus  by (\ref{Cov}) and (\ref{Cov1}) we obtain
\begin{align*}
&\frac{b}{n}\mathrm{Cov}\{\gamma(z_1),\gamma(z_2)\}=E\{\Sigma(z_1,z_2)\}\\&=
\frac{\partial^2}{\partial z_1\partial z_2}\Big(-2\int \log\big(1-g(z_1)g(z_2)\hat u(k)\big)dk-g(z_1)g(z_2)u(0)
+\kappa_4g^2(z_1)g^2(z_2)\Big)+o(1)\\
&=:\mathcal{C}(z_1,z_2)+o(1),
\end{align*}
where we used also that by (\ref{eq_3}) $\zeta^{-1}=g(z_1)g(z_2)$.
According to the definition (\ref{N(phi)}) and the above relation
\begin{align*}
&\frac{b}{n}\mathrm{Var}\{\mathcal{N}_n[\varphi_\eta]\}\to\int d\lambda_1d\lambda_2\varphi(\lambda_1)\varphi(\lambda_1)
C_\eta(\lambda_1,\lambda_2),
\end{align*}
where
\begin{align*}
C_\eta(\lambda_1,\lambda_2)=&\frac{1}{4\pi^2}\big(\mathcal{C}(\lambda_1+i\eta,\lambda_2-i\eta)+
\mathcal{C}(\lambda_1-i\eta,\lambda_2+i\eta)\\
&-\mathcal{C}(\lambda_1+i\eta,\lambda_2+i\eta)-\mathcal{C}(\lambda_1-i\eta,\lambda_2-i\eta)\big).
\end{align*}
Now by Proposition \ref{p:CLTcont} for any $\varphi$ possessing the norm (\ref{norm}) we have
\begin{align*}
\frac{b}{n}\mathrm{Var}\{\mathcal{N}_n[\varphi]\}\to\lim_{\eta\to 0}
\int d\lambda_1d\lambda_2\varphi(\lambda_1)\varphi(\lambda_1)
C_\eta(\lambda_1,\lambda_2).
\end{align*}
Let us make the change of variables $\lambda_1=2\cos x_1$, $\lambda_2=2\cos x_2$.
Then, using that (see (\ref{g}))
\[\lim_{\eta\to+ 0}g(\lambda_\alpha\pm i\eta)=-e^{\mp ix_\alpha},\quad\alpha=1,2,
\]
we obtain (\ref{var}) by a simple calculus.

\section{Auxiliary results}

\textit{Proof of Lemma \ref{l:b_var}}

The first identity of (\ref{repr}) yields that  it suffices to estimate
$E\{|A'_pA^{-1}_p-E_{1}\{A'_pA^{-1}\}|^2\}$. Note that for any $a$ independent of
$\{w_{1i}\}$  we have
\[E_p\{|\xi^\circ_p|^2\}\le E_p\{|\xi-a|^2\}. \]
Hence it suffices to estimate
\[\bigg|\frac{A'_p}{A_p}-\frac{E_{p}\{A'_p\}}{E_{1}\{A_p\}}\bigg|=
\bigg|\frac{A'^\circ_p}{E_{p}\{A\}}-
\frac{A'^{\circ}_p}{E_{1}\{A_p\}}\,\frac{A'_p}{A_p}\bigg|\le
\bigg|\frac{A'^{\circ}_p}{E_{p}\{A_p\}}\bigg|+
\bigg|\frac{A^\circ_p}{yE_{p}\{A_p\}}\bigg|.
\]
Here and below $z=x+iy$, $y>0$. Let us use also the relation (\ref{sp_rel})
%\begin{equation}\label{sp_rel}
%\Im(G^{(1)}m^{(1)},m^{(1)})=\Im z(|G^{(1)}|^2m^{(1)},m^{(1)}),\quad \Im\hbox{Tr }G^{(1)}=\Im z\hbox{Tr }|G^{(1)}|^2.
%\end{equation}
which  yields, in particular, that $|{A'_p}/{A_p}|\le y^{-1}$ .
Using (\ref{A,Q}), we get

\begin{equation}\label{b_v.2b}
E_p\Big\{\Big|\frac{A^\circ_p}{E_{p}\{A_p\}}\Big|^2\Big\}\le
\frac{Cb^{-2}\hbox{Tr }G^{(p)}I_pG^{(p)}*}{|E_{p}\{A_p\}|^2},
\end{equation}
 Similarly
\[E_p\bigg\{\bigg|\frac{A'^\circ_p}{E_{p}\{A_p\}}\bigg|^2\bigg\}\le
\frac{Cb^{-2}\hbox{Tr }(G^{(p)})^2I_p(G^{(p)*})^2}{|E_{p}\{A_p\}|^2}\le\frac{Cb^{-2}\hbox{Tr
}G^{(p)}I_pG^{(p)*}}{|E_{p}\{A_p\}|^2}.
\]
Thus
\begin{align}\label{b.*}\frac{b}{n}E\{|(\gamma_n(z))^\circ|^2\}\le Cn^{-1} \sum_p\frac{\hbox{Tr
}G^{(p)}I_pG^{(p)*}}{by^2|E_{p}\{A_p\}|^2}.\end{align}
Notice that  the H\"{o}lder inequality implies for any $\delta>0$
\begin{align*}&\int\Big|b^{-1}\sum_{|j-p|\le bC} G^{(p)}_{jj}(x+iy)\Big|^{1+\delta}dx\le
Cb^{-1}\sum_{|j-p|\le bC}\int|G^{(p)}_{jj}(x+iy)|^{1+\delta}dx\\
&\le
b^{-1}\sum_{j}\int\sum_k\frac{|(\psi_k,e_j)|^2}{|(x-\lambda_k)^2+y^2|^{(1+\delta)/2}}dx\le C\delta^{-1} y^{-\delta}.
 \end{align*}
 Hence, denoting $\mathcal{L}_p=\{x:|\sum u_{pj}G^{(p)}_{jj}(x+iy)|>1\}$, we obtain
 \[\int 1_{\mathcal{L}_p}dx\le C\min_{\delta}\{\delta^{-1} y^{-\delta}\}\le C\log y^{-1}.\]
Then, using once more that by (\ref{sp_rel}) each summand in  the r.h.s. of (\ref{b.*}) is bounded by $y^{-4}$,
we get
\begin{align*}&\int\frac{b}{n}E\{|(\gamma_n(zx+iy))^\circ|^2\}dx\\
&\le Cn^{-1}\sum_{p}\Big(\int_{\mathbb{R}\setminus([-1,1]\cup\mathcal{L}_p)}(y^2b)^{-1}\hbox{Tr }G(x+iy)I_pG^{(p)*}(x+iy)dx
+Cy^{-4}\int_{[-1,1]\cup\mathcal{L}_p}dx\Big)\\
&\hskip8cm\le Cy^3+Cy^{-4}\log y^{-1} \le C'y^{-4}\log y^{-1}.
 \end{align*}

$\square$

\textit{Proof of Lemma \ref{l:G_ii}}.
It follows from (\ref{repr}) that
\begin{align}\label{G_ii.0}
E\{|G_{pp}-E\{G_{pp}\}|^2\}&\le|\Im z|^{-2}E\{|A_p-E\{A_p\}|^2\}\\
&\le 2|\Im z|^{-1}(E\{|\bar A_p-E\{\bar A_p\}|^2\}
+\mathrm{Var}\{ A_p^\circ\})\notag\\&
\le 2|\Im z|^{-2}b^{-1}\sum u_{pi}\mathrm{Var}\{G^{(p)}_{ii}\}+Cb^{-1}.
\notag\end{align}
But since
\begin{align*}
E\{|G_{ii}-G^{(p)}_{ii}|\}\le|\Im z|^{-2}b^{-1}E\{|(G^{(p)}v^{(p)})_i|^2\}\le C|\Im z|^{-2}b^{-1},
\end{align*}
we have
\begin{align}\label{G_ii.1}
\mathrm{Var}\{G^{(p)}_{ii}\}= \mathrm{Var}\{G_{ii}\}+O(b^{-1})=\mathrm{Var}\{G_{pp}\}+O(b^{-1}).
\end{align}
Here the last equality is due to the invariance of the distribution of $M$ with respect to the "shift" $i\to (i+1)\mod(n)$.
Hence for any $z:|\Im z|\ge 2$ we obtain from (\ref{G_ii.0})
\begin{align}\label{G_ii.2}
\mathrm{Var}\{G_{pp}\}\le2|\Im z|^{-2}\mathrm{Var}\{G_{pp}\}+Cb^{-1}\mathrm{Var}\{G_{pp}\}\le 2Cb^{-1}.
\end{align}
Let us fix any $z=x+i\eta$ with $0<\eta<2$ and consider the function
\[\phi(\zeta)=\log (c_0b^{1/2}|\mathrm{Cov}\{G_{pp}(\zeta),G_{pp}(z)\}|) \]
in the half-circle $\Omega=\{\Im \zeta<2\}\cap\{|\zeta-x-2i|\le |2-\eta/2|\}$. It is a harmonic function, and
in view of (\ref{G_ii.2}) for $\Im \zeta=2$ we can choose $c_0$ sufficiently small to have
\begin{align*}
c_0b^{1/2}|\mathrm{Cov}\{G_{pp}(\zeta),G_{pp}(z)\}|\le
c_0 b^{1/2}\mathrm{Var}^{1/2}\{G_{pp}(\zeta)\}\mathrm{Var}^{1/2}\{G_{pp}(z)\}\le 1\\
\Rightarrow \phi(\zeta)\le 0,\quad \zeta\in\gamma_1:=\partial\Omega\cap \{\Im \zeta=2\}.
\end{align*}
Moreover, in view of the trivial bound $|G_{pp}(\zeta)|\le|\Im \zeta|^{-1}$, we have
\[
\phi(\zeta)\le\log b^{1/2}+\log c_0\eta^{-2},\quad \zeta\in\gamma_2:=\partial\Omega\cap\{|\zeta-x-2i|= |2-\eta/2|\}.
\]
Hence, by the theorem on two constants (see \cite{Evgr}, p. 296), we have
\begin{align}\label{G_ii.3}
\phi(\zeta)\le(\log b^{1/2}+\log c_0\eta^{-2})\omega(\zeta),
\end{align}
where the harmonic function
\[\omega(\zeta):=\frac{2}{\pi}\Im\log\frac{2-\eta/2-(\zeta-x-2i)}{2-\eta/2+(\zeta-x-2i)},\]
satisfy the conditions
\[\omega(\zeta)=0,\;\zeta\in\gamma_1,\qquad\omega(\zeta)=1,\;\zeta\in\gamma_2.\]
Since $\omega(z)=1-2\delta$ with some $\delta(\eta)>0$, (\ref{G_ii.3}) implies the first line of  (\ref{var_G}):
\[c_0b^{1/2}\mathrm{Var}\{G_{pp}(z)\}\le(c_0b^{1/2})^{1-2\delta}\Rightarrow \mathrm{Var}\{G_{pp}(z)\}\le Cb^{-\delta}.\]
Using (\ref{repr}), (\ref{G_ii.0}), and (\ref{G_ii.3}), we get similarly to(\ref{G_ii.1}),
\begin{align*}E\{G_{pp}(z)\}=-(z+E\{G_{pp}(z)\})^{-1}+O(b^{-\delta})\\
\Rightarrow E\{G_{pp}(z)\}=
g(z)+O(b^{-\delta}).\end{align*}
 Thus, we have proved the second line of (\ref{var_G}).

$\square$


\begin{thebibliography}{99}
\small

\bibitem{Ba-Si:04} Z.Bai, J.W.Silverstein.  CLT for linear spectral statistics
of large-dimensional sample covariance matrices. Ann.Probab, \textbf{ 32},  553-605 (2004)

\bibitem{Bi:95} P. Billingsley. Probability and measure. Wiley Series in Probability and Mathematical Statistics. John
Wiley  Sons Inc., New York, third edition, 1995. A Wiley-Interscience Publication.

\bibitem{EK:11} L.Erd\"{o}s,  A.Knowles. Quantum diffusion and eigenfunction delocalization in a
random band matrix model. Comm. Math. Phys. \textbf{303}, 509 – 554 (2011)

\bibitem{EKY:13} L.Erd\"{o}s,  A.Knowles,  H.-T.Yau, J.Yin. Delocalization and diffusion profile for
random band matrices. Electron. J. Probab. \textbf{18} n 59, 1–58 (2013)

\bibitem{Evgr} M.A.Evgrafov : Analytic Functions. Dover Pubns (1978)

\bibitem{FM:91} Y.V.Fyodorov, A.D.Mirlin: Scaling properties of localization in random band
matrices: a у-model approach, Phys. Rev. Lett. \textbf{67}, 2405 – 2409 (1991)

\bibitem{Gu:13} F.Benaych-Georges, A.Guionnet, C.Male. Central limit theorems
for linear statistics of heavy tailed random matrices
 Comm. Math. Phys. \textbf{329}, n 2, 641–686 (2014)

\bibitem{JSSo:14} I.Jana, K.Saha, A.Soshnikov.
 Fluctuations of Linear Eigenvalue Statistics of Random Band Matrices
 arXiv:1412.2445

\bibitem{KK:02}  A.Khorunzhy, W.Kirsch. On asymptotic expansions and scales of spectral universality in
band random matrix ensembles, Comm. Math. Phys. \textbf{231}, 223-255 (2002).

\bibitem{MKP:92} S.A.Molchanov, L.A.Pastur,  A.M.Khorunzhy. Eigenvalue distribution for band random matrices
in the limit of their infinite rank. Teor. Matem. Fizika \textbf{90}, 108 – 118 (1992)

\bibitem{Li-So:13} L.Li, A. Soshnikov.
Central Limit Theorem for Linear Statistics of Eigenvalues of Band Random Matrices
Random Matrices: Theory and Applications, \textbf{2}, 04 (2013)

\bibitem{LP:09}   A. Lytova, L. Pastur. Central limit theorem for linear eigenvalue statistics of random matrices
with independent entries, Annals of Probability \textbf{ 37}, n5, 1778-1840 (2009)

\bibitem{NY:14} J.Najim, J.Yao. Gaussian fluctuations for linear spectral statistics of large random covariance matrices
 arXiv:1309.3728

\bibitem{PS:11} L.Pastur, M.Shcherbina. Eigenvalue Distribution of Large Random Matrices. Mathematical Survives and Monographs,
V171, American Mathematical Society: Providence, Rhode Island (2011)

\bibitem{ORSo:14} S.O'Rourke, D.Renfrew, A.Soshnikov, V.Vu. Products of independent elliptic random matrices
 arXiv:1403.6080

\bibitem{S:11} M.Shcherbina. Central Limit Theorem for linear eigenvalue statistics of the Wigner
 and sample covariance random matrices. Journal of Mathematical Physics, Analysis, Geometry
V7, N2, pp 176-192, 2011.

\bibitem{ST:12} M.Shcherbina, B.Tirozzi.
 Central limit theorem for fluctuations of linear eigenvalue statistics of large random graphs. Diluted regime.
  arXiv:1111.5492[math-ph]. J.Math.Phys. \textbf{53}, 1-18  (2012)

\bibitem{TS:12} T.Shcherbina. On the second mixed moment of the characteristic polynomials of the 1D band matrices
Commun. Math. Phys. \textbf{328}, 1-28 (2012).

\bibitem{Sp}  T.Spencer: SUSY statistical mechanics and random band matrices. Lecture notes
in mathematics 2051 (CIME Foundation subseries), Quantum many body system

\bibitem{W} E.P.Wigner. On the distribution of the roots of
certain symmetric matrices. Ann.Math. \textbf{67},  325-327 (1958)

\end{thebibliography}
\end{document}